\documentclass[twocolumn,english,aps,prl,superscriptaddress]{revtex4-1}

\usepackage{bm}
\usepackage{amsmath}
\usepackage{amssymb}
\usepackage{graphicx}



\begin{document}

\author{Dmitry Solenov}

\email{d.solenov@gmail.com}

\address{Naval Research Laboratory, Washington, District of Columbia 20375,
USA}

\author{Kirill A. Velizhanin}

\email{kirill@lanl.gov}

\address{Theoretical Division and Center for Nonlinear Studies, Los Alamos
National Laboratory, Los Alamos, NM 87545, USA}

\title{Adsorbate Transport on Graphene by Electromigration}
\begin{abstract}
Chemical functionalization of graphene holds promise for various applications
ranging from nanoelectronics to catalysis, drug delivery, and nanoassembly.
In many applications it is important to be able to transport adsorbates
on graphene in real time. We propose to use electromigration to drive
the adsorbate transport across the graphene sheet. To assess the efficiency
of electromigration, we develop a tight-binding model of electromigration
of an adsorbate on graphene and obtain simple analytical expressions
for different contributions to the electromigration force. Using experimentally
accessible parameters of realistic graphene-based devices as well
as electronic structure theory calculations to parametrize the developed
model, we argue that electromigration on graphene can be efficient.
As an example, we show that the drift velocity of atomic oxygen covalently
bound to graphene can reach $\sim$1 cm/s. 
\end{abstract}

\maketitle

Many unique properties of graphene -- a monoatomic crystalline sheet
of carbon -- stem from the fact that it ``lacks'' volume and is,
therefore, a truly 2d ``all-surface'' material \cite{Geim2007-183}.
For instance, the surface functionalization of graphene (e.g., graphene
oxide) provides an opportunity to alter electronic properties of the
entire material, which holds promise in nanoelectronics \cite{Boukhvalov2008-4373,Englert2011-279},
nonvolatile memory \cite{Cui2011-6826}, graphene-based nanoassemblies
for catalysis, photovoltaics and fuel cells applications \cite{Kamat2010-520,Kamat2011-242}.
In majority of these applications, the performance of a graphene-based
device can be significantly improved {\em provided} there is a
way to tune the surface functionalization in real time, i.e., during
device operation. The related problem is to control and direct the
transport of adsorbed atoms/molecules for nanoassembly \cite{Yang2008-124709}
and drug delivery applications \cite{Voloshina2011-220}.

Electromigration is the drift of material on the surface (or in the
bulk) of a current-carrying conductor \cite{Sorbello1997-159}. We
propose to exploit electromigration as an efficient and easily controllable
method to drive the directed transport of adsorbates on graphene.
Very recently, and for the first time, the efficient electromigration
of metallic clusters/atoms on graphene has been demonstrated experimentally
\cite{Barreiro2011-775}. In the current work, however, we focus on a different
class of adsorbates (perhaps more relevant from the perspective of
chemical functionalization) -- atoms or molecules {\em covalently}
bound to graphene. At first glance, the covalent binding may seem
too strong to allow for efficient electromigration. Nevertheless,
recent theoretical studies suggest the possibility of fast diffusion.
Specifically, the activation energy for the atomic oxygen diffusion
has been shown to drop from $\sim$0.7~eV for neutral graphene to
$\sim$0.15~eV for $n$-doped graphene, resulting in a diffusion
coefficient as high as $\sim10^{-6}$~cm$^{2}$/s \cite{Suarez2011-146802}.

\begin{figure}
\includegraphics[width=0.99\columnwidth]{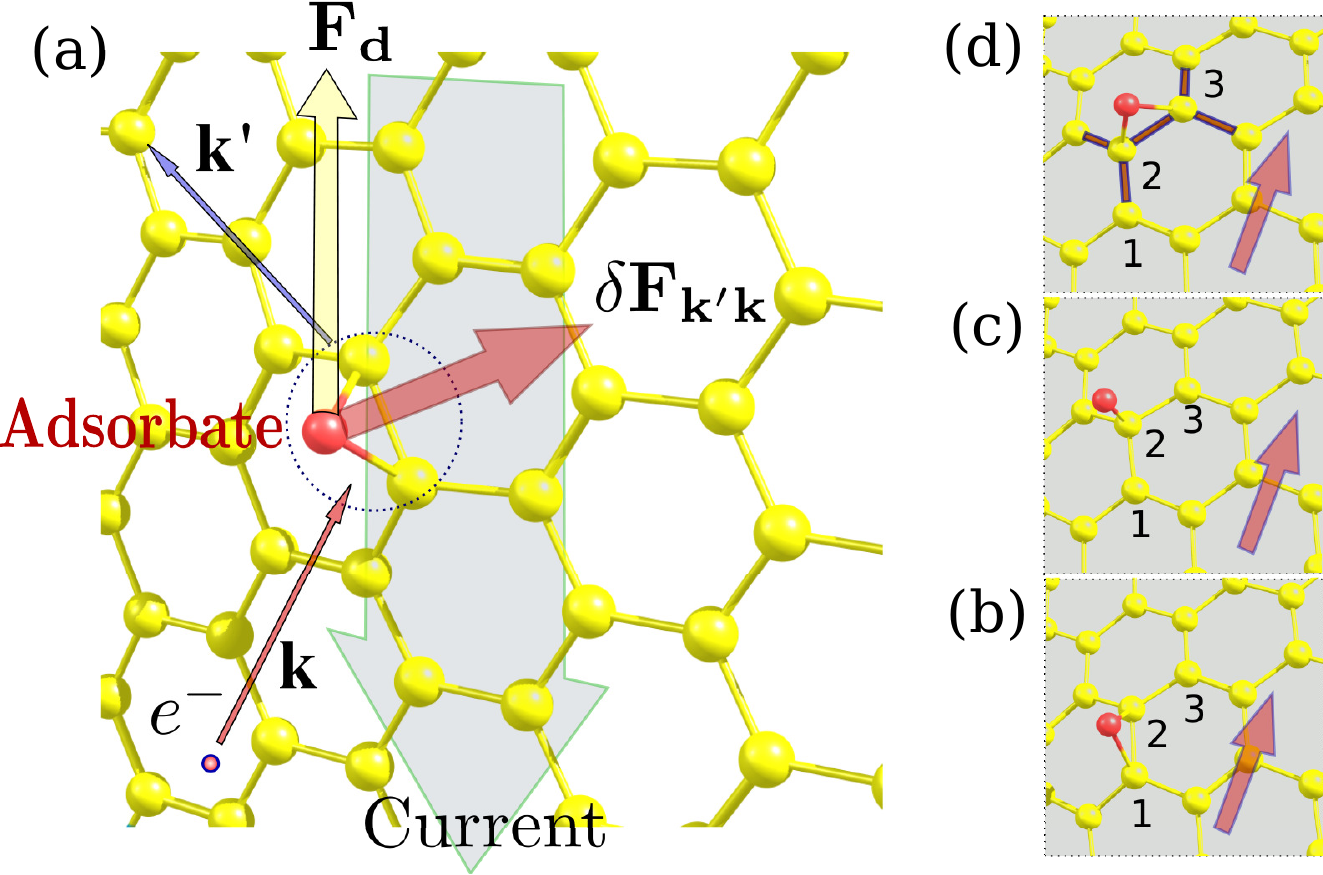}

\caption{\label{fig:Schematic} Electromigration on graphene. (a) Schematic
depiction of the electron electromigration force on an adsorbate (oxygen
atom) bound to graphene surface: direct force ${\bf F}_{{\bf d}}$
and the force due to an electron scattering event, $\delta{\bf F}_{{\bf k}',{\bf k}}$, contributing to
electron wind force ${\bf F}_{w}$. The adsorbate is negatively charged
and both ${\bf F}_{{\bf d}}$ and ${\bf F}_{w}$ are directed opposite to the current. The current flows downward,
as shown. (b),(c),(d) Adsorbate hopping: the adsorbate remains bound
while hopping (e.g., from position 1-2 to position 2-3). The arrow
indicates the net drift direction. The highlighted bonds in inset
(d) are excluded from the tight-binding description due to the presence
of a covalently bound adsorbate (see text).}
\end{figure}

In this Letter, we investigate electromigration of an adsorbate covalently
bound to graphene, focusing on atomic oxygen (O) and the amino group (NH),
as an example. In the lowest energy configuration, a single oxygen
atom covalently binds to two adjacent carbon atoms of the graphene's
honeycomb lattice, thus forming an epoxy bond, Figs.~\ref{fig:Schematic}(a)-\ref{fig:Schematic}(d). 

A single nitrogen atom can bind similarly, providing its third covalent
bond for chemical functionalization, e.g., in drug delivery applications.
In what follows, we develop a simple tight-binding model and obtain
an analytical result for the drift velocity of an adsorbate as a function
of its charge, the electric current in graphene, as well as the backgate
doping level, and temperature. Specifically, we find that the migration
(drift) velocity reaches up to $\sim0.6-4$ cm/s at electrical current
densities of $\sim1$ A/mm and temperatures of $300-500$ K. The doping
level and temperature dependence of the drift velocity are then used
to formulate a robust adsorbate manipulation technique based on local heating of a graphene sample.
This technique can become suitable for patterning and other applications involving direct access
to the surface of graphene.

\begin{figure}
\includegraphics[width=0.6\columnwidth]{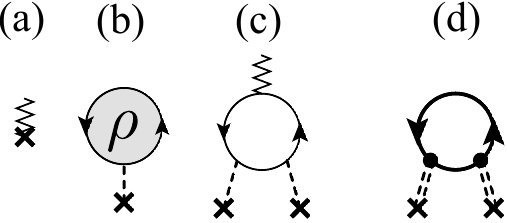}

\caption{\label{fig:Diagrams} Diagrams representing various contributions
to the electromigration force (see text for details). Zigzag lines
denote the external dc electric field, crosses stand for the
adsorbate scattering vertexes. Dashed lines are the adsorbate-substrate
interaction. The grey-filled blob denotes the electron density in graphene. 
Solid lines stand for the equilibrium propagator of electrons in graphene.
Thick solid lines denote fully dressed nonequilibrium electron propagators
(in the presence of the electric field). Double dashed lines represent the screened
adsorbate-substrate interaction.}
\end{figure}

A particle (or defect) in contact with a conductor experiences an
electromigration force which can be written as \cite{Sorbello1997-159}
\begin{equation}
{\bf F}={\bf F}_{d}+{\bf F}_{w}=eZ{\bf E}-\langle\bm{\nabla}_{{\bf R}}\hat{U}\rangle,\label{eq:TotForce}
\end{equation}
where $e$, $Z$ and ${\bf E}$ are the absolute value of the electron
charge, the charge of the particle in atomic units, and the vector of the
external dc electric field (EF), respectively. The first term, the
direct force ${\bf F}_{d}$, originates from the direct interaction
of the adsorbate's charge with the EF, as shown diagrammatically in Fig.~\ref{fig:Diagrams}(a). 

The second term, ${\bf F}_{w}=-\langle\bm{\nabla_{{\bf R}}}\hat{U}\rangle$ [Fig.~\ref{fig:Diagrams}(b)], originates from the scattering of electrons in graphene by the adsorbate-graphene interaction potential, $\hat{U}$
(${\bf R}$ is the position of the adsorbate). This contribution
is often referred to as the {\em electron wind force}. Assuming the adsorbate
{\em structureless}, i.e, treating $U$ as just a single-particle
scattering potential, this contribution can be rewritten as 
\begin{equation}
{\bf F}_{w} = \sum_{{\bf k},{\bf k}'}\delta{\bf F}_{{\bf k}',{\bf k}} =\frac{i}{\hbar}\sum_{{\bf k},{\bf k}'}\hbar({\bf k}'-{\bf k})U_{{\bf k}'{\bf k}}\rho_{{\bf k}{\bf k}'},\label{eq:wind}
\end{equation}
where $\rho_{{\bf k}{\bf k}'}=\langle\hat{c}_{{\bf k}'}^{\dagger}\hat{c}_{{\bf k}}\rangle$
is the single-particle density matrix and $\hat{c}_{{\bf k}}^{\dagger}$
($\hat{c}_{{\bf k}}$) creates (annihilates) a Bloch wave with quasimomentum
${\bf k}$. Equation~(\ref{eq:wind}) has a very appealing microscopic
interpretation: it describes electron scattering from ${\bf k}$-state
to ${\bf k}'$-state within graphene, accompanied by $\hbar({\bf k}-{\bf k}')$
momentum transfer to the adsorbate, with amplitude $U_{{\bf k}'{\bf k}}\rho_{{\bf k}{\bf k}'}$,
see Fig.~\ref{fig:Schematic}(a). The pitfall is that Eq.~\ref{eq:wind}
is exact only for free electrons, and it can be nontrivial to justify
that the {\em momentum} transfer is equal to the change of {\em quasimomentum}
in the presence of band-structure effects \cite{Sorbello1997-159}.
In what follows, we will show that the dominant contribution to
$U$ is sufficiently smooth (as compared to the size of the graphene
unit cell), so that Eq.~(\ref{eq:wind}) holds.

To obtain a closed expression for ${\bf F}_{w}$, the density matrix
$\rho_{{\bf k}{\bf k}'}$ in Eq.\ (\ref{eq:wind}) is expanded up
to the leading order in $U$. The validity of such perturbative expansion
will be justified later when we address the graphene-specific form
of $U$. The first-order contribution in $U$ to ${\bf F}_{w}$ vanishes exactly. The simplest nonvanishing
contribution (second order in $U$, first order in ${\bf E}$)
describes the interaction of an adsorbate with the current-carrying
charge density of graphene, Fig.~\ref{fig:Diagrams}(c).
Important processes not described by this diagram include (i) Coulomb-induced screening
of $U$ and (ii) scattering due to impurities and phonons.
Most diagrams containing such processes, as well as the effect of
the EF, can be ``lumped together'' by dressing the adsorbate-graphene
interaction using the random phase approximation and replacing bare electron
propagators with fully dressed nonequilibrium (i.e., current-carrying)
Green's functions. As a result, the electron wind contribution to
the electromigration force, depicted diagrammatically in Fig.~\ref{fig:Diagrams}(d),
can be expressed as 
\begin{equation}
{\bf F}_{w}=-\sum_{{\bf k},{\bf k}'}\hbar({\bf k}'-{\bf k})\delta\Gamma_{{\bf k}'{\bf k}},\label{eq:F}
\end{equation}
\begin{equation}
\delta\Gamma_{{\bf k}'{\bf k}}=\frac{2\pi}{\hbar^2}|\tilde{U}_{{\bf k}'{\bf k}}|^{2}[1-f_{j}({\bf k}')]f_{j}({\bf k})\delta(\omega_{k'k}),\label{eq:Gm}
\end{equation}
where $\tilde{U}_{{\bf k}'{\bf k}}$ is the screened adsorbate-graphene
potential. The steady-state distribution function in graphene is given by 
\begin{equation}
f_{j}({\bf k})=f_{0}(k)-\frac{4\pi({\bf j}\cdot{\bf k})}{ev_{F}k_{F}^{2}}\delta(k-k_{F}),\label{eq:FDcurrent}
\end{equation}
in the linear response approximation \cite{Ashcroft1976}. Here ${\bf j}$
is the current density within graphene and $f_{0}(k)$ is the Fermi-Dirac
distribution; the Fermi momentum and the velocity of Dirac electrons in
graphene are denoted by $k_{F}$ and $v_{F}$, respectively. Equations
(\ref{eq:F}) and (\ref{eq:Gm}) agree with the general (second-order)
result for the wind force obtained earlier in Ref.~\cite{Sorbello1997-159}.
To complete the derivations we need to obtain $\tilde{U}_{{\bf k}{\bf k}'}$,
which is specific to the case of adsorbates on graphene.

The interaction potential between an adsorbate and graphene has
two distinct contributions: $\tilde{U}=\tilde{U}^{C}+\tilde{U}^{{\rm def}}$.
The first one, $\tilde{U}^{C}$, describes the scattering of electrons in
graphene by the Coulomb potential of the charged adsorbate.
The second one, $\tilde{U}^{{\rm def}}$, describes the scattering due
to the lattice defect caused by the covalent bonding of an adsorbate to graphene. As will become clear shortly, the Coulomb contribution is dominant and, therefore, we discuss it
first.

In the backgated graphene with experimentally accessible electron
densities, the screened Coulomb potential varies slowly over graphene
unit cell, and it suffices to consider scattering only within a single
Dirac cone. Within the lowest energy tight-binding description and
the random phase approximation for screening, the Coulomb contribution
can be cast in the form \cite{Rana2007-155431,CastroNeto2009-109} 
\begin{equation}
\tilde{U}_{{\bf k}'{\bf k}}^{C}=\frac{2\pi e^{2}Z/\tilde{\kappa}}{|{\bf k}'-{\bf k}|+k_{TF}}\frac{(1+e^{i\theta_{{\bf kk'}}})}{2}e^{-|{\bf k}'-{\bf k}|h},\label{eq:Coulomb}
\end{equation}
where $k_{TF}=4e^{2}k_F/(\hbar\tilde{\kappa}v_F)\approx9k_F/\tilde{\kappa}$
is the Thomas-Fermi momentum. For graphene laid on top of a half-space
dielectric substrate with the dielectric constant $\kappa$, the effective
constant is given by $\tilde{\kappa}=(\kappa+1)/2$ \cite{Ponomarenko2009-206603}.
In what follows, a SiO$_{2}$ substrate with $\tilde{\kappa}=2.5$ will be assumed.
The angle $\theta_{{\bf kk'}}$ stands for the angle
between the vectors ${\bf k}$ and ${\bf k}'$, both measured with respect
to the same Dirac point. The distance between the adsorbate and graphene,
$h$, is comparable to the graphene lattice constant, $a$, and, hence,
$|{\bf k}'-{\bf k}|h\leq2k_{F}h\ll1$ is negligible.

The effective size of the potential $\tilde{U}^{C}$ is given by $k_{TF}^{-1}$,
so that at $\mu=0.2$ eV (easily reached in backgated graphene) one has
$k_{F}a\approx0.08$, and, therefore, $k_{TF}a\approx0.3$. Therefore,
the size of the potential is significantly larger than the unit cell,
justifying both Eq.~(\ref{eq:F}) and the neglect of scattering between
the Dirac cones. 

Using the inequality $k_{TF}\gg k_{F}$, the screened
Coulomb potential of the adsorbate simplifies to $\tilde{U}_{{\bf k}'{\bf k}}^{C}\approx \pi e^{2}Z(1+e^{i\theta_{{\bf kk'}}})/\tilde{\kappa}k_{TF}$.
Substituting this expression, along with Eq.~(\ref{eq:FDcurrent}),
into Eq.~(\ref{eq:Gm}) and retaining only terms linear in ${\bf j}$,
we finally obtain for the Coulomb contribution to the electron wind
force \cite{Note1}
\begin{equation}
{\bf F}_{w}^{C}=-\frac{\hbar{\bf j}}{e}\left(\frac{\pi Z}{4}\right)^{2}.\label{eq:wforce}
\end{equation}
Here, the spin and valley degeneracies are lumped into the current ${\bf j}$.

To prove that the Coulomb contribution to the electron wind force
is dominant, we estimate the amplitudes of Coulomb and lattice defect
contributions. The covalent binding of an adsorbate to graphene alters
the hybridization of the involved carbon atoms from $sp^{2}$ to $sp^{3}$, see
Figs.~\ref{fig:Schematic}(b)-\ref{fig:Schematic}(d). This amounts to cutting out the corresponding
$p_{z}$ carbon orbitals from the tight-binding description of graphene,
creating a lattice point defect. The lack of a single $p_{z}$ orbital
can be mimicked by a fictitious impurity potential which cancels out
the hopping integrals involving this orbital. For electrons near the
Fermi circle such an impurity potential is essentially a delta function
with a magnitude in momentum space of $U^{{\rm def}}\sim ta^{2}\sim\hbar v_{F}a$,
where $t\approx$~2.8~eV is the hopping energy. Comparing $U^{C}$
and $U^{{\rm def}}$ (screened or unscreened), we obtain 
\begin{equation}
U^{{\rm def}}/U^{C}(k_{F})\approx\frac{\hbar v_{F}}{e^{2}}\frac{1}{2\pi|Z|}k_{F}a\approx\frac{0.07}{|Z|}k_{F}a.
\end{equation}
Since $k_{F}a\ll1$ (see above), the contribution of the single-bond
defect potential is small compared to that of the Coulomb potential
at not too small $Z$. The actual defect potential leads to exclusion
of several bonds, e.g., five for oxygen in the equilibrium state;
see Fig.~\ref{fig:Schematic}(d). Furthermore, the defect potential
cannot be considered smooth on the scale of the unit cell, which might
introduce significant band-structure effects [see the discussion following
Eq.~(\ref{eq:wind})]. Nevertheless, our analysis (not provided)
using the accurate tight-binding model showed that these two complications
do not change the qualitative conclusion that we have drawn: the $U^{{\rm def}}$
contribution to the electron wind force is small compared to the Coulomb
one and will be omitted henceforth.

Finally we note that the second-order representation of the wind force,
Eq. (\ref{eq:Gm}) and Fig.~\ref{fig:Diagrams}(d), is sufficient
for adsorbates with not very large $Z$. As will be seen shortly,
$|Z|\leq$0.4 for both oxygen and nitrogen, and, therefore, the screened
Coulomb potential is sufficiently small relative to the Fermi energy
$\tilde{U}^{C}(k_{F})k_{F}^{2}/v_{F}k_{F}\approx|Z|/2\leq0.2$.
At larger $Z$, multiple events of adsorbate-induced scattering of in-graphene electrons
can modify the results obtained here.

\begin{figure}
\includegraphics[bb=0bp 0bp 800bp 472bp,width=0.99\columnwidth]{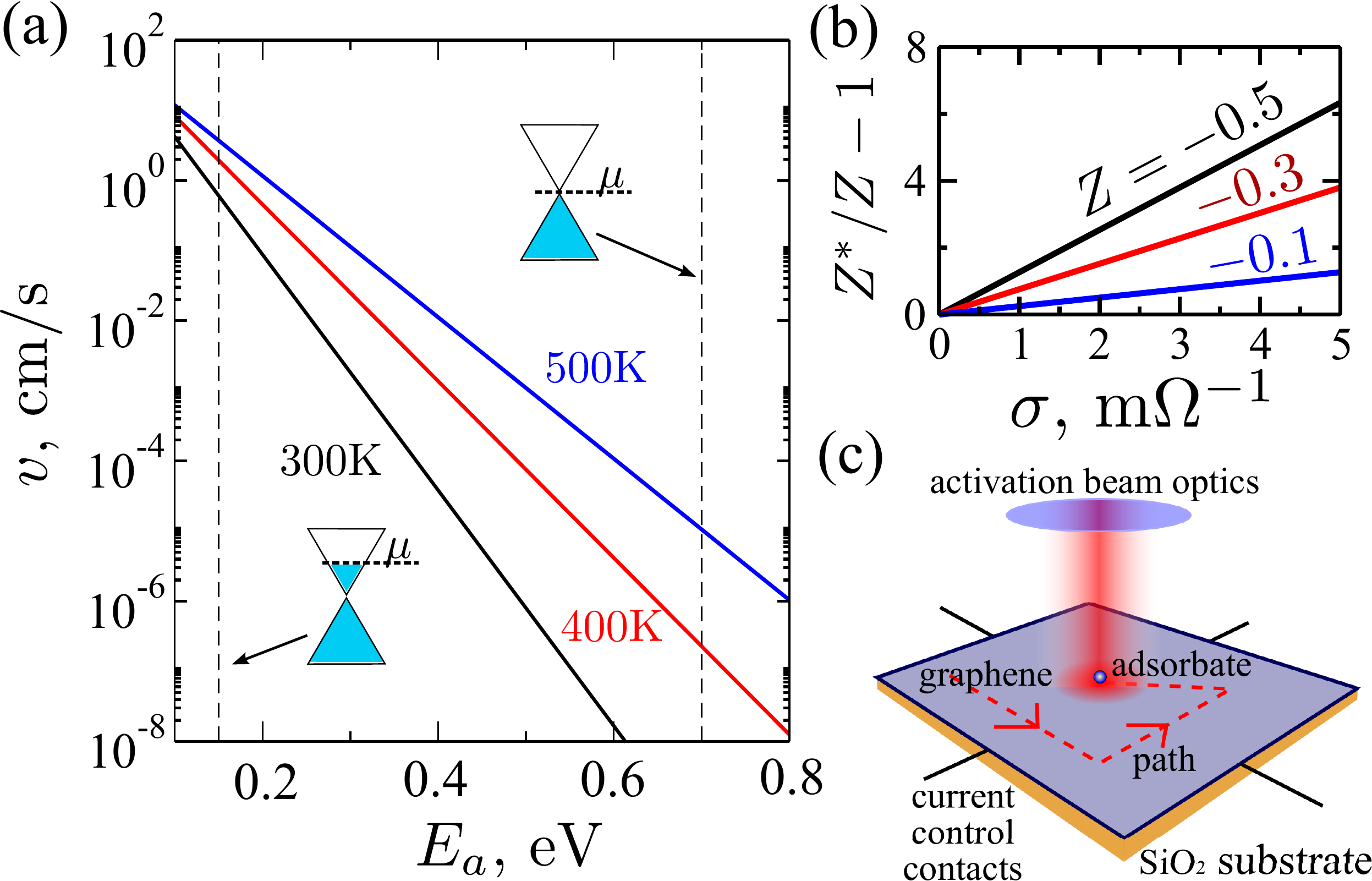}

\caption{\label{fig:mrate} Drift of an adsorbate along the graphene sheet. (a)
Drift velocity, $v$, for $Z$=-0.3 $j=1$~A/mm, $\sigma=1$~m$\Omega^{-1}$
and different temperatures. Dashed vertical lines mark activation
energies of 0.15~eV and 0.7~eV, which correspond to oxygen diffusion
on $n$-doped and charge-neutral graphene, respectively \cite{Suarez2011-146802}.
(b) Relative contribution of the electron wind force to the effective
charge of the adsorbate. (c) Robust control of activation and drift
path with a local heater, e.g. laser beam.}
\end{figure}

The total driving force of the electromigration for a charged adsorbate
on graphene is given by 
\begin{equation}
{\bf F}={\bf F}_{d}+{\bf F}_{w}=eZ{\bf j}/\sigma-\frac{\hbar{\bf j}}{e}\left(\frac{\pi Z}{4}\right)^{2}=eZ^{*}{\bf j}/\sigma,\label{eq:fforce}
\end{equation}
where the {\em effective} charge of the adsorbate is introduced
as $Z^{*}=Z-\frac{\sigma Z^{2}}{8\sigma_{0}}$ and $\sigma_{0}=\frac{2e^{2}}{\hbar\pi^{2}}$
is the minimal conductivity of graphene \cite{CastroNeto2009-109}.
Equation~(\ref{eq:fforce}) -- the main result of this work -- expresses
the total driving force of the electromigration via experimentally accessible
parameters such as conductivity of graphene and the current density.
The charge of the adsorbate, however, is not directly accessible, so we
estimated it by performing electronic structure theory
calculations using Gaussian~09 quantum chemistry package \cite{Gaussian09}.
These calculations were performed using density functional theory
with the PW91 functional \cite{Perdew1996-16533} and the 6-311G**
basis set for a number of graphene flakes of increasing size
(up to ${\rm C_{62}}$) to guarantee
the convergence with respect to boundary effects. A single adsorbate was put in the center of a flake.
The charges of the adsorbates
were found in the range $Z_{{\rm O}}\in(-0.2,-0.4)$ and $Z_{{\rm NH}}\in(-0.1,-0.3)$,
depending on the specific position of the adsorbate along the hopping
trajectory [see Fig.~\ref{fig:Schematic}(b)-\ref{fig:Schematic}(d)], and the graphene doping
level. The ratio of the electron wind and direct forces (or, equivalently,
$Z^{*}/Z-1$) is shown in Fig.~\ref{fig:mrate}(b) (as a function
of graphene conductivity \cite{Novoselov2005-197}) for the range
of $Z$ values. 
As is seen, the electron wind
contribution dominates the electromigration force except for smallest $Z$ or low conductivity,.

To calculate the drift velocity of an adsorbate on graphene due to
electromigration, we use the Einstein--Smoluchowski relation between
the diffusion coefficient, $D$, and the drift velocity, ${\bf v}$,
i.e., 
\begin{equation}
{\bf v}={\bf F}D/k_{B}T,\quad\quad D=\frac{d^{2}\nu_{0}}{4}e^{-E_{a}/k_{B}T},\label{eq:DiffCoeff}
\end{equation}
where $d=1.23$~\AA{}~is the hopping distance. The attempt frequency
for oxygen was found to be $\nu_{0}=26$~THz in Ref.~\cite{Suarez2011-146802}.
We assume the same attempt frequency for the NH group, having in mind
the approximate character of the calculations and the fact that the diffusion
coefficient is not overly sensitive to variations of this parameter
(compared to, e.g., the activation energy or temperature). The activation
energy for oxygen diffusion, $E_{a}$, as obtained in Ref.~\cite{Suarez2011-146802},
is $\sim$0.7~eV for charge-neutral graphene, and a much lower value
of $\sim$0.15 eV for $n$-doped graphene with the charge density
of $-7.6\times10^{13}$~cm$^{-2}$. Our electronic structure theory
calculations qualitatively confirm the strong sensitivity of the activation
energy for oxygen diffusion. Furthermore, we found a similar dependence
of the activation energy on doping level for NH adsorbate, albeit with
somewhat higher activation energies (by $0.2-0.3$~eV).
Figure~\ref{fig:mrate}(a) shows the dependence of the drift velocity
on $E_{a}$ for $Z$=-0.3 at $j$=1A/mm and $\sigma$=10$^{-3}$ $\Omega^{-1}$.
Specifically, at these parameters and $E_{a}$=0.15~eV the drift
velocity is $\sim$6~mm/s at room temperature ($T$=300~K) and reaches
up to 4~cm/s at $T$=500~K. 

The increased drift velocity of the electromigration at higher temperatures
can be used to perform adsorbate manipulations and patterning via
a guided motion of adsorbates along the surface of graphene. This
is most easily achieved by heating graphene locally with a focused
laser beam [Fig.~\ref{fig:mrate}(c)], or by a heated AFM tip \cite{Wei2010-1373}.
This local heating will enhance $v$, while keeping the drift velocity
low outside the heating spot by adjusting the doping level to lower
values. As a result, it should be possible to move adsorbates along
the desired path and assemble them into desired patterns by tracing
the motion with the ``local heater'' and choosing the appropriate
current directions via a set of source-drain contacts in perpendicular
directions (typical, e.g., for Hall conductance measurements), see
Fig~\ref{fig:mrate}(c). The spatial resolution of patterning is
thermodynamically limited by the balance between diffusion and drift,
and is given by $l_{p}=kT/F$ \cite{vanKampen1992}. For the current
density and conductivity used in Fig~\ref{fig:mrate}(a) we obtain
$l_{p}=0.09T/Z^{*}$ nm K$^{-1}$. For adsorbed oxygen atoms at $T=300$~K,
the result in $l_{p}\approx 50$ nm. 

An experimental verification of an efficient electromigration of adsorbates
on graphene can either be done using AFM/STM techniques to directly
monitor diffusion/drift of adsorbates, or by optical means, e.g.,
adopting a nitrogen-based adsorbate with a fluorescent functional group
and monitoring the fluorescence of such adsorbates with temporal and
spatial resolution. Mapping a trajectory of adsorbates on carbon materials
with finite band gap (e.g., carbon nanotubes) by photoluminescence 
{\em quenching} is an alternative strategy \cite{Crochet2012-126}. 

We thank Sergei Tretiak and Sergei Ivanov for consulting on electronic
structure theory calculations. We also acknowledge valuable discussions
with Ivar Martin, Tom Reinecke, and Chad Junkermeier. This work was
performed under the NNSA of the U.S. DOE at LANL under Contract No.
DE-AC52-06NA25396, and, in part, by ONR and NAS/LPS.


\end{document}